\documentclass[onecolumn,floatfix,showpacs,nofootinbib,11pt,groupedaddress,superscriptaddress,]{revtex4-1}
\usepackage{amssymb,amsmath,graphicx}
\usepackage{color}
\usepackage[colorlinks,hyperindex]{hyperref}
\hypersetup{hidelinks,backref=true,pagebackref=true,hyperindex=true,colorlinks=true,breaklinks=true,urlcolor= blue}
\hypersetup{%
  colorlinks = true,
  linkcolor  = blue
}
\usepackage{bookmark}
\usepackage{hyperref,color,xcolor}
\hypersetup{hidelinks,backref=true,pagebackref=true,hyperindex=true,colorlinks=true,breaklinks=true,urlcolor= blue}
\hypersetup{%
  colorlinks = true,
  linkcolor  = blue
}
\usepackage{makeidx}
\newcommand{\beq}{\begin{eqnarray}}
\newcommand{\benu}{\begin{enumerate}}
\newcommand{\enu}{\end{enumerate}}
\newcommand{\eeq}{\end{eqnarray}}
\newcommand{\lp}{\ell_{\rm p}}
\newcommand{\mpl}{m_{\rm p}}
\newcommand{\gn}{G_{\rm N}}
\renewcommand{\d}{{\rm d}}
\newcommand{\be}{\begin{equation}}
\newcommand{\ee}{\end{equation}}

\newcommand{\ba}{\begin{eqnarray}}
\newcommand{\ea}{\end{eqnarray}}
\begin{document}
\title{GUP Hawking fermions from MGD black holes}
\author{Roberto Casadio}
\email{casadio@bo.infn.it}
\affiliation{Dipartimento di Fisica e Astronomia, Universit\`a di Bologna, 
via Irnerio 46, 40126 Bologna, Italy}
\affiliation{INFN, 
Sezione di Bologna, viale B.~Pichat 6, 40127 Bologna, Italy}
\author{Piero Nicolini}
\email{nicolini@fias.uni-frankfurt.de}
\affiliation{Frankfurt Institute of Advanced Studies (FIAS) and Institut f\"ur Theoretische Physik, Goethe Universitat,
Frankfurt am Main, Germany}
\author{Rold\~ao~da~Rocha}
\email{roldao.rocha@ufabc.edu.br}
\affiliation{Centro de Matem\'atica, Computa\c c\~ao e Cogni\c c\~ao, Universidade Federal do ABC, 09210-580,
Santo Andr\'e, Brazil.}
\begin{abstract}
We derive the Hawking spectrum of fermions emitted by a minimally geometric deformed (MGD) black hole.
The MGD naturally describes quantum effects on the geometry in the form of a length scale related, for instance,
to the existence of extra dimensions.
The dynamics of the emitted fermions is described in the context of the generalised uncertainty principle (GUP)
and likewise contains a length scale associated with the quantum nature of space-time.
We then show that the emission is practically indistinguishable from the Hawking thermal spectrum for large black hole
masses, but the total flux can vanish for small and finite black hole mass.
This suggests the possible existence of black hole remnants with a mass determined by the two length scales.
\end{abstract}
\maketitle
\section{Introduction}
The minimal geometric deformation (MGD) was originally proposed~\cite{ovalle2007,ovalle2008} as a systematic
method to determine high-energy corrections to general relativistic (GR) spherically symmetric solutions in the
brane-world~\cite{ovalle2008,covalle2,covalle3,darkstars}.
It was also used to study bulk effects on realistic stellar interiors~\cite{Ovalle:2013xla} and the hydrodynamics of
black strings~\cite{Casadio:2013uma}. 
Recently, the MGD corrections to the gravitational lensing was estimated in Ref.~\cite{Cavalcanti:2016mbe},
and it was shown that the merging of MGD stars could be detected by the eLISA/LIGO experiments
more easily than their Schwarzschild counterparts~\cite{daRocha:2017cxu}. 
Finally, it was proposed that the MGD can be realised in analogue gravitational systems which can be studied
in laboratories~\cite{daRocha:2017lqj}.
In particular, in brane-world models~\cite{maartens}, our Universe is a (codimension-1) brane with tension
$\sigma$ and the MGD leads to a deformation of the  Schwarzschild metric proportional to a positive length
scale $\ell\sim\sigma^{-1}$.
Quite interestingly, the MGD was also shown to apply to more general departures from GR than those predicted within
the extra-dimensional scenario~\cite{Ovalle:2017fgl}, being stable under small linear perturbations~\cite{Fernandes-Silva:2017nec}.
In the following, we shall therefore consider the scale $\ell\sim\lp$ as related to a generic departure from GR induced
by quantum physics.
\par
The most renown quantum effect that should occur around black holes is the Hawking evaporation~\cite{hawking}.
There are many derivations of this effect, most of which just assume a classical background geometry.
One of such approaches is the tunnelling method~\cite{2,3,6,Vanzo:2011wq}, which has been considered 
both for bosons and fermions~\cite{k0,k1} in various types of black hole backgrounds.
The (closely related) WKB approximations have then been employed in order to calculate quantum corrections to
the Bekenstein-Hawking entropy, {\em e.g.}~for the Schwarzschild black hole.  
The tunnelling method was recently employed in order to compute the Hawking radiation spectrum due to dark
spinors~\cite{hoff,cav}.
Concerning in particular the emission of spin-1/2 fermions, the Hawking radiation was analysed as the tunnelling of
Dirac particles through an event horizon, where quantum corrections in the single particle action are proportional
to the usual semiclassical contribution. 
The effects of the spin of each type of spin-1/2 fermions were then shown to cancel out, due to the isotropy of the emission.
\par 
We shall here study the Hawking radiation of fermions from MGD black holes, including the quantum effects
on the fermion dynamics predicted by the Generalised Uncertainty Principle (GUP), which has been
comprehensively explored in Refs.~\cite{gup1,gup12,gup13,gup14,gup2,gup21,gup22,gup23,gup3,scardigli,gup4,gup51,gup5,SC},
being compatible with a unitary description.
 Such effects are also characterised by a (minimum) length scale $\beta\sim\lp^2$, and can lead either to the formation of hot \cite{gup5} or cold \cite{IMN13} black hole remnants, or sub-Planckian black holes \cite{CaMN15}. 
It is therefore interesting to compare the changes to the quantum dynamics encoded by $\beta$ with the changes related to the (quantum) MGD represented by $\ell$.
Although Refs.~\cite{Mu:2015qta,an1,Chen:2013tha} already analysed the GUP effects, no associated MGD effects
have been studied in this context.
Hence, we shall here present a more general picture that can cover each of the above cases, by simply tuning the respectively
relevant parameters. 
In particular, we shall compute the corrected Hawking flux by means of the tunnelling method, and show that
the quantum modifications essentially depend on $\ell$, with $\beta$ generating only sub-leading deviations
(at least according to this approximation). The choice of spinors as emitted particles is based on the fact that the Hawking radiation favors lower spin, lighter particles and it is expected to be dominated by fermions \cite{Page:1976df}. As a result we can rely on the formalism developed in ~\cite{hoff,cav}.
\par
The paper is organised as follows:
in the next Section, we will briefly review the MGD black hole metric for which 
the tunnelling rate for fermions will be computed in Section~\ref{SIII};
conclusions and comments are then summarised in Section~\ref{SIV}.
\section{The MGD black hole}
\label{MGD}
We recall that, in the brane-world scenario, the Gauss-Codazzi projection of the five-dimensional Einstein equations
yields the effective four-dimensional Einstein equations~\cite{maartens},
\be
\label{5d4d}
R_{\mu\nu}-\frac12\, R\,g_{\mu\nu}
=
8\,\pi\,\gn\,T^{\rm eff}_{\mu\nu}
-\Lambda\,g_{\mu\nu}
\ ,
\ee
where $\gn= \lp/\mpl$, with $\mpl$ and $\lp$ the four-dimensional Planck mass and scale, respectively;
$R_{\mu\nu}$ and $R$ are the Ricci tensor and scalar of the four-dimensional metric; 
$\Lambda$ is the cosmological constant (which we shall neglect hereafter). 
The effective stress tensor in Eq.~(\ref{5d4d}) contains the matter energy-momentum tensor on the brane,
the electric component of the Weyl tensor and the projection of the bulk energy-momentum tensor onto the
brane~\cite{maartens}. 
For static and spherically symmetric metrics,  
\be
\label{abr}
\d s^{2}
=
-A(r)\, \d t^{2} + \frac{\d r^{2}}{B(r)}
+r^{2}\left(\d\vartheta^2+\sin^2\vartheta\, \d\varphi^2\right)
\ ,
\ee
the MGD provides a solution to Eqs.~\eqref{5d4d} by deforming the radial metric component of the corresponding
GR solution~\cite{covalle2,darkstars}.  
For the GR Schwarzschild metric, and dismissing terms of order $\sigma^{-2}$ or higher, one obtains~\cite{covalle2}
\begin{subequations}
\ba
\label{nu}
A(r)
&=&
1-\frac{2\,\gn\,M}{r}
\ ,
\\
B(r)
&=&
A(r)\left[1+\frac{2\,{\ell}}{{2\,r-{3\,\gn\,M}}}\right]
\ ,
\label{mu}
\ea
\end{subequations}
where $\ell\sim \sigma^{-1}$ is the length scale previously discussed in the Introduction and $M$ the ADM
mass.
There are two solutions for the equation $B(r)=0$, namely 
\begin{subequations}
\ba
r_+
&=&
2\,\gn\,M\ ,
\\
r_-
&=&
\frac{3\,\gn\,M}{2}-\ell
=
\frac{3}{4}\,r_+-\ell
\ ,
\ea
\end{subequations}
so that $r_+>r_-$ for any $\ell>0$.
For studying the Hawking radiation, we are interested in the region outside $r_+$, that effectively acts as the event horizon,
and just note that $r_-$ is not a (Cauchy) horizon~\cite{covalle2}. 
\par
We just mention in passing that an explicit expression for $\ell$ in terms of $\sigma^{-1}$ can be obtained
by first considering a compact source of finite size $r_0$ and proper mass $M_0$~\cite{covalle2,ovalle2007},
and then letting the radius $r_0$ decrease below $r_+$.
However, for practical purposes, it is more convenient and general to show the dependence on the length $\ell$,
as we mentioned in the Introduction.
For example, observational data impose bounds on the length $\ell$, from which bounds on $\sigma$ can be
straightforwardly inferred according to the underlying model~\cite{Casadio:2015jva,Casadio:2016aum}.  
\section{Hawking flux for fermions}
\label{SIII}
Let us now consider a GUP in the form~\footnote{In our units, $\hbar=\lp\,\mpl$.}
\be
\Delta x\, \Delta p
\gtrsim
\frac{\hbar}{2}\left[1+ \beta\, \Delta p^2\right]
\ ,
\ee
where $\beta =\beta_0/\mpl^2$, and $\beta_0 $ is a (dimensionless) parameter encoding 
quantum gravity effects on the particle dynamics.
The upper bound $\beta_0 < 10^{21}$ was recently obtained (see~\cite{Das:2008kaa,Scardigli:2014qka}
and references therein).
In this framework, the position and momentum operators are respectively given by $x_i = X_{i}$
and $p_i = P_i\,(1 + \beta\, p^2)$, with $i=1,2,3$.
The variables $X_{i}$ and $P_{i}$ then satisfy the canonical commutation
relations $\left[X_{j},P_{k}\right]= i\, \hbar\, \delta_{ij}$, yielding
\be
p^2
=
-\hbar^2\,g_{ij}\,\partial^i\, \partial^j
\left(1 - 2\,\beta\, \hbar^2\, g_{kl}\,\partial^k\partial^l\right)
\ .
\label{eq2.3}
\ee
The generalised frequency is defined by $\tilde \omega = E\,( 1 - \beta\, E^2)$
for $ E = i\, \hbar\,\partial_t$.
On the mass shell, the energy of a particle with mass $m$ and electric charge $e$
reads~\cite{Chen:2013tha,ACS}
\be
E
=\tilde E
\left[
1 + \beta\left(p^2 + m^2\right)
\right]
\ . 
\label{eq2.5}
\ee
The Dirac equation in an external electromagnetic field ${\rm A}^\mu$ (with $\mu=0,\ldots,3$)
is given by
\be
\left\{
i\,\gamma^{\mu}
\left[\hbar\left(\partial_{\mu}+\Omega_{\mu}\right)
+i\,e\,{\rm A}_{\mu}\right]
+m
\right\}
\psi
=0
\ ,
\label{eq2.6}
\ee
where $\Omega _\mu \equiv\frac{i}{2}\,\omega _\mu^{\,\alpha\beta}\,\Sigma_{\alpha\beta}$, and
$\omega _\mu^{\,\alpha\beta}=-e^\rho_{\,\beta}\partial_\mu e_\rho^{\,\alpha}+e_\nu^{\,\alpha} e^\rho_{\,\beta} \Gamma^\nu_{\mu\rho}$
is the spin connection.
Eqs.~(\ref{eq2.3}) and (\ref{eq2.5}) can be replaced into Eq.~(\ref{eq2.6}), yielding the equation~\cite{Chen:2013tha}
\be
\left\{i\,\hbar\,\gamma^{0}\partial_{0}
+
\left[m-e\,\gamma^{\mu}\,{\rm A}_{\mu}
+i\,\hbar\,\gamma^{\mu}
\left(\Omega_{\mu}+\hbar\,\beta\,\partial_\mu
\right)
\right]
\left(
1-
\beta\,m^{2}
+\beta\,\hbar^{2}\,g_{jk}\,\partial^{j}\,\partial^{k}
\right)
\right\}
\psi
=
0
\ .
\label{eq2.9}
\ee
\par
\begin{figure}[htb!]
\begin{minipage}{14pc}
\includegraphics[width=18pc]{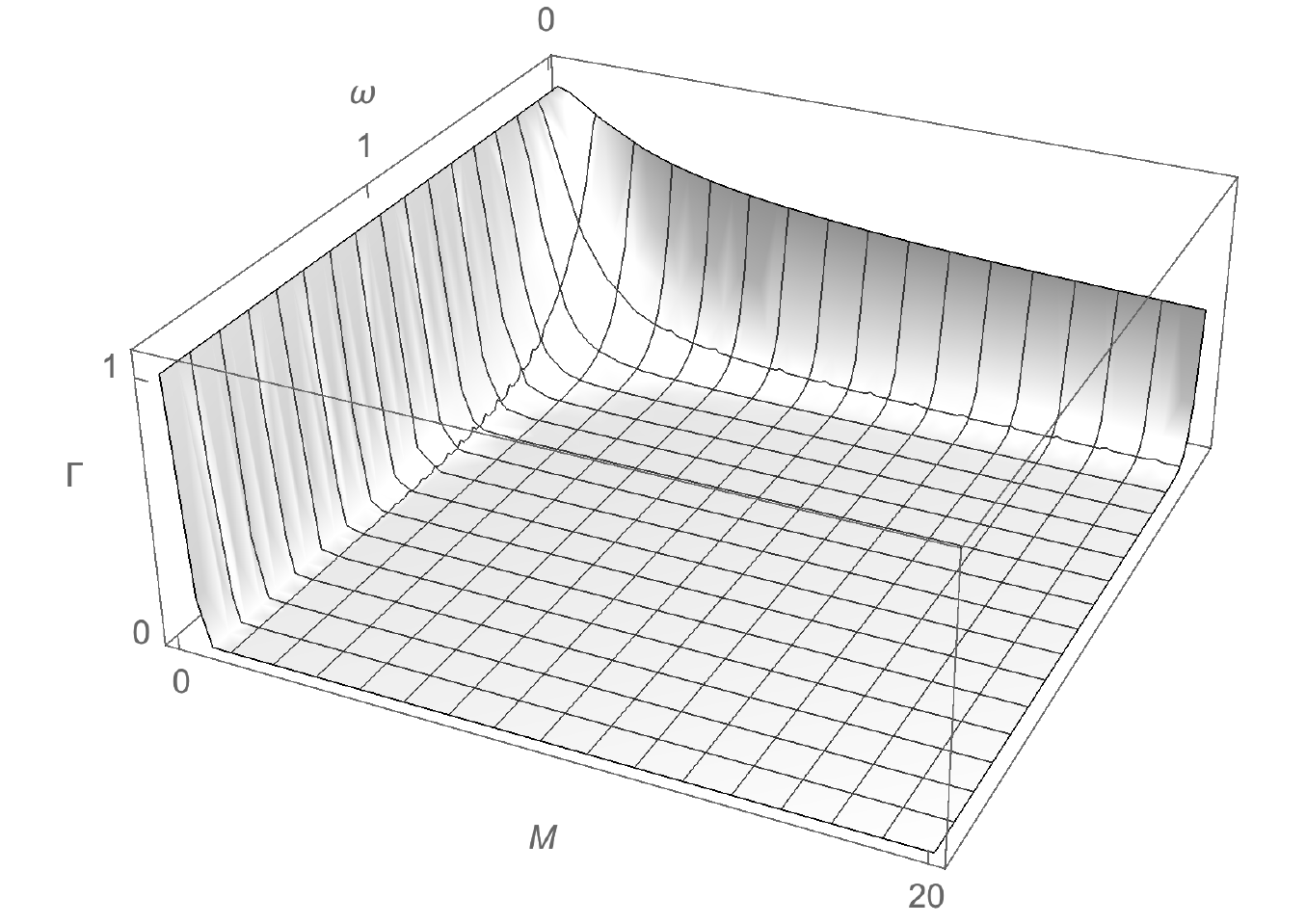}
\caption{\label{b1}\footnotesize{Spectrum of the Hawking radiation, as a function of the frequency $\omega$ and black hole mass $M$ (in powers of $\mpl$), for $\beta_0=10^7$ and $\ell_0=10^{-10}$.}}
\end{minipage}\hspace{5pc}
\begin{minipage}{14pc}
\includegraphics[width=17.5pc]{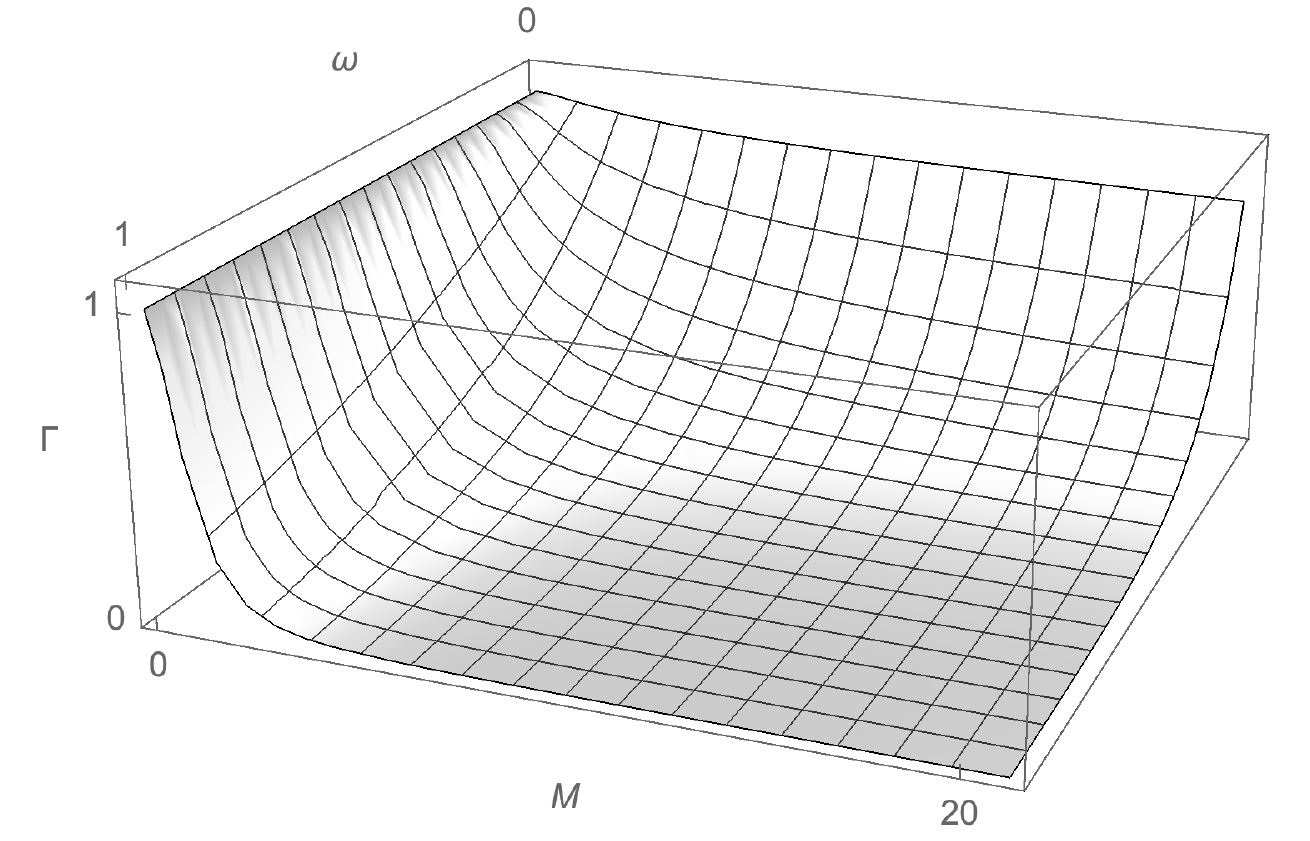}
\caption{\label{b2}\footnotesize{Spectrum of the Hawking radiation, as a function of the frequency $\omega$ and black hole mass $M$ (in powers of $\mpl$),
for $\beta_0=0=\ell_0$.
}}
\end{minipage}
\begin{minipage}{14pc}
\includegraphics[width=16pc]{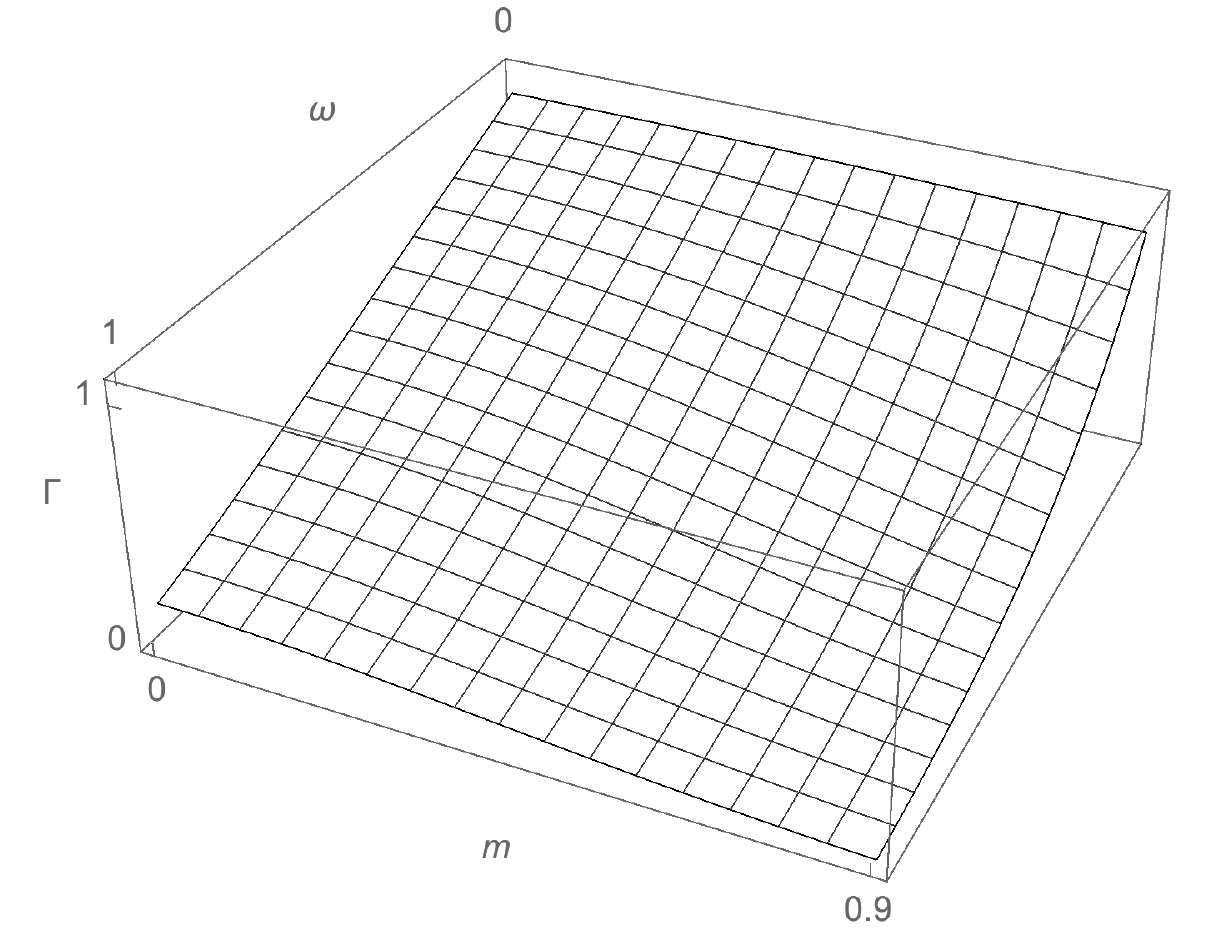}
\caption{\label{b3}\footnotesize{Boltzmann factor, as a function of the frequency $\omega$ and fermion mass $m$, for $\beta_0=10^7$ and $\ell_0=10^{-10}$.}}
\end{minipage}\hspace{5pc}
\begin{minipage}{14pc}
\includegraphics[width=15.5pc]{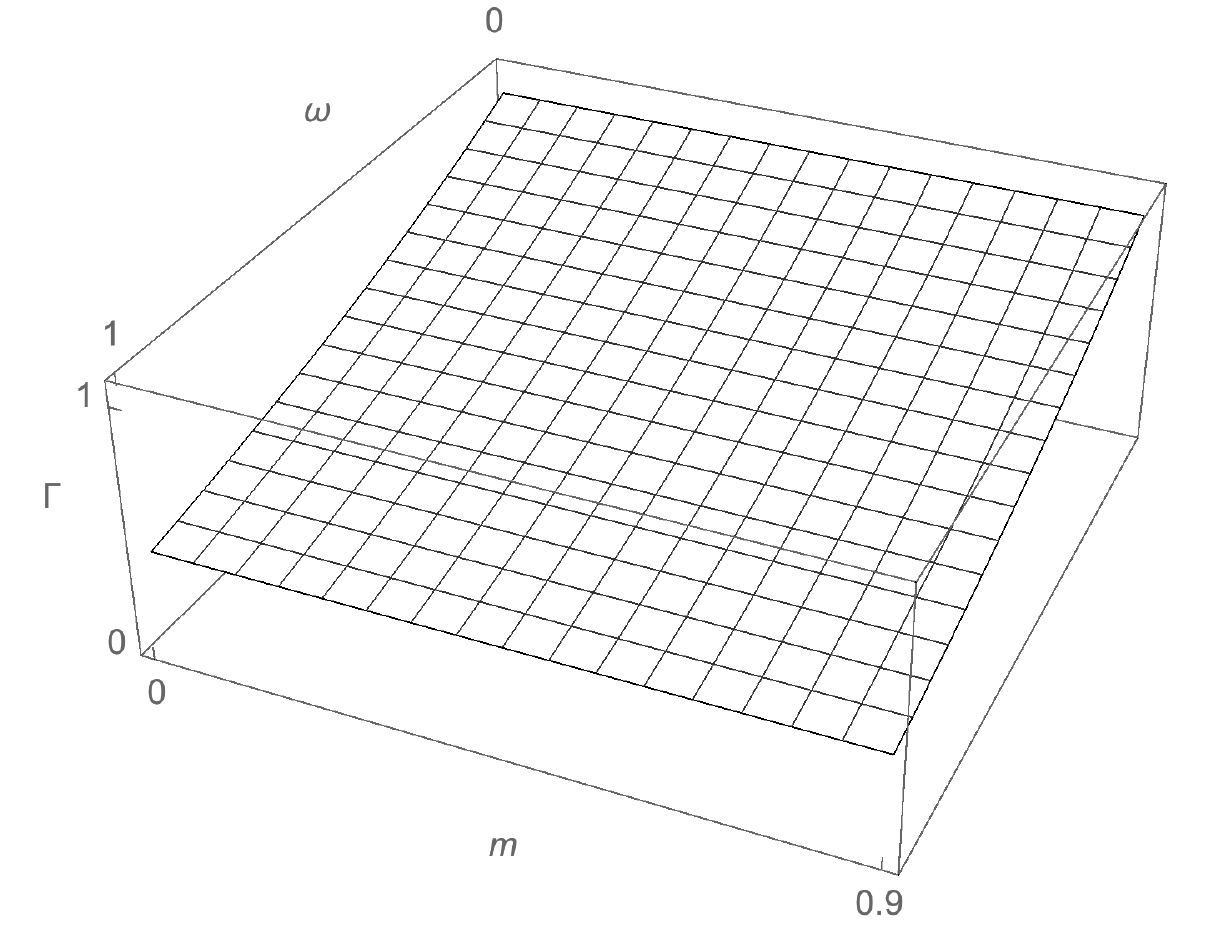}
\caption{\label{b4}\footnotesize{Spectrum of the Hawking radiation, as a function of the frequency $\omega$ and
fermion mass $m$, for $\beta_0=0=\ell_0$.
}}
\end{minipage}
\includegraphics[width=17pc]{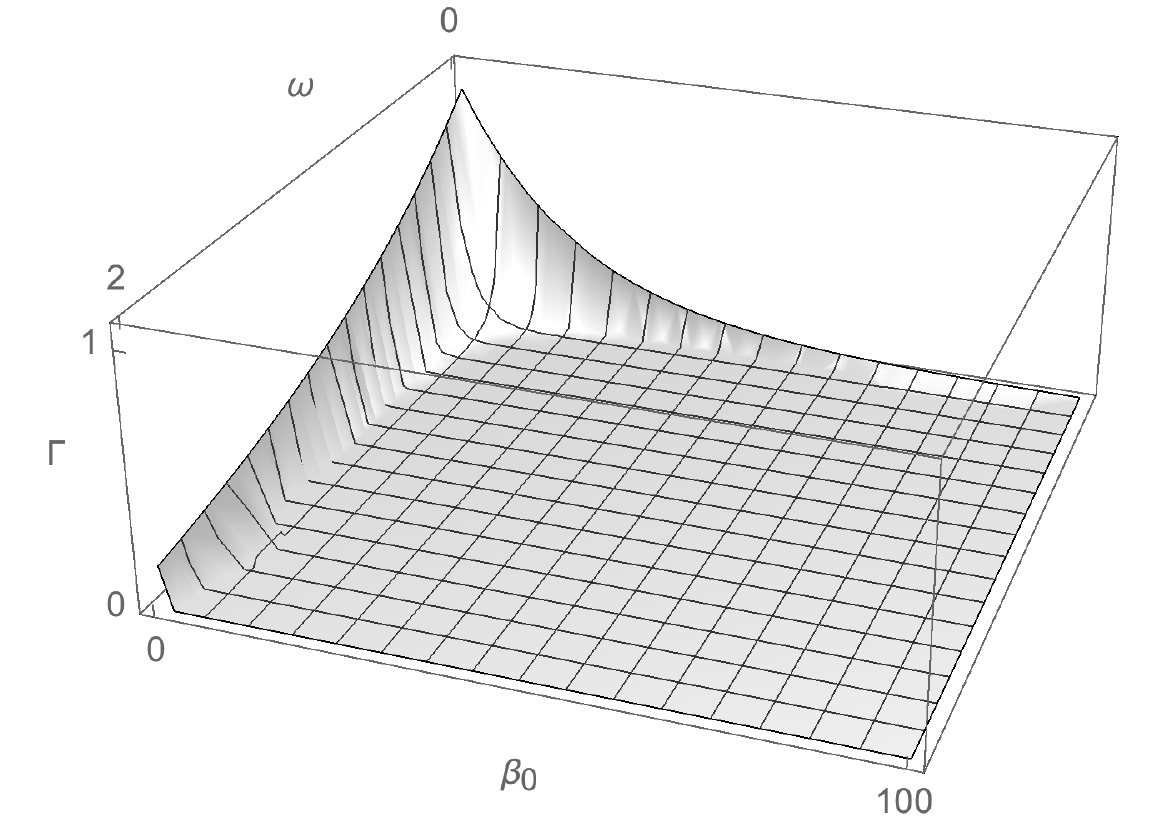}
\caption{\label{bs3}\footnotesize{Spectrum of the Hawking radiation, as a function of the frequency $\omega$ and
GUP parameter $\beta_0$, for $\ell_0=10^{-10}$. 
}}
\end{figure}

The Hawking radiation emitted by black holes can contain several kinds of particles.
Hereon, we shall analyse the tunnelling of regular fermions across the event horizon of the MGD
black hole~\eqref{nu}-\eqref{mu}.
The regular spinor field describing the fermion is assumed to be~\cite{hoff,cav}
\be
\Psi
=
\left(
\psi,0,\mathring\psi,0\right)^\intercal\,
\exp\!\left\{\frac{i}{\hbar}\,I(t,r,\theta,\phi)\right\}
\ ,
\label{304}
\ee
for an action $I$ and wave-functions ${\psi}$ and ${{\mathring\psi} }$.
The metric~(\ref{abr}) yields the tetrads
\be
e_\mu^{\,\alpha}
=
{\rm{diag}}
\left[
\sqrt{A(r)}, \frac{1}{\sqrt{B(r)}}, r, r\,\sin\theta
\right]
\ ,
\ee
and the $\gamma^\mu$ matrices read 
\ba
&&
\gamma^{t}
=
\frac{1}{\sqrt{A(r)}}
\left(
\begin{smallmatrix}
i & 0&0&0\\
0 &i&0&0\\
0&0& -i&0\\
0&0&0&-i
\end{smallmatrix}
\right)
\ ,
\qquad\qquad
\gamma^{\theta}
=
\frac{1}{r}
\left(
\begin{smallmatrix}
0 &0&0&1\\
0&0&1&0\\
0&1&0&0\\
1&0&0&0
\end{smallmatrix}\right)
\ ,
\nonumber
\\
&&
\gamma^{r}
=
\sqrt{B(r)}
\left(
\begin{smallmatrix}
0 &0&1&0\\
0&0&0&-1\\
1& 0&0&0\\
0&-1&0&0
\end{smallmatrix}\right)
\ ,
\qquad\qquad
\gamma^{\phi}
=
\frac{1}{r\sin\theta}
\left(
\begin{smallmatrix}
0 &0&0&-i\\
0&0&i&0\\
0&-i&0&0\\
i&0&0&0
\end{smallmatrix}\right)
\ .
\label{305}
\ea
Inserting Eqs.~(\ref{304}) and~\eqref{305} into the Dirac equation~(\ref{eq2.9}), 
the WKB approximation to leading order in $\hbar$ yields the equations of motion
\ba
&&
{\psi}
\left\{
\frac{i}{\sqrt{A}}
\left[
\partial_{t}I-e\,{\rm A}_t\left(1-\beta\, m^{2}-\beta\, K\,\right)
\right]
-m
\left(1-\beta\, m^{2}+\beta\,K\right)
\right\}
\nonumber
\\
&&
=
{\mathring\psi}
\left(
1-\beta\,m^{2}+\beta\,K
\right)
\sqrt{B}\,\partial_{r}I
\quad
\label{306}
\\
&&
{\mathring\psi}
\left\{
\frac{i}{\sqrt{A}}
\left[\partial_{t}I
+e\,{\rm A}_t
\left(1-\beta\, m^{2}-\beta\, K\right)
\right]
+m
\left(1-\beta\, m^{2}+\beta\, K\right)
\right\}
\nonumber
\\
&&
=
-\psi\left(1-\beta\,m^{2}+\beta\,K\right)
\sqrt{B}\,\partial_{r}I
\quad
\label{307}
\\
&&
\left(1-\beta\,m^2-\beta\,K\right)
\left(
\partial_\theta I
+i\,\frac{\partial _ {\phi}I}{\sin\theta}
\right)\label{388}
=
0
\ ,
\ea
with 
\be
K
=
B\,(\partial _r I)^2
+
\frac{(\partial _ {\theta}I)^2}{r^2}
+
\frac{(\partial _ {\phi}I)^2}{r^2\sin^2\theta}
\ .
\ee 
Upon writing the action in the usual form 
\be
I
=
-\omega\, t
+W(r)
+{\mathit{\Theta}}(\theta , \phi)
\ ,
\label{3010}
\ee
where $\omega$ is the energy of the emitted fermions, the tunnelling probability can now be
derived~\cite{2,Vanzo:2011wq,an1}.
Inserting Eq.~(\ref{3010}) into 
Eq.~(\ref{388}), one obtains
\be
\left(
\frac{\partial _ {\phi}{\mathit{\Theta}}}{\sin\theta}
-i\,{\partial _ {\theta}{\mathit{\Theta}}}
\right)
\left[
\beta B\,(W')^2
+
\beta \frac{(\partial _{\theta}{\mathit{\Theta}})^2}{r^2}
+
\beta \frac{(\partial _ {\phi}{\mathit{\Theta}})^2}{r^2\,\sin^2\theta}
+\beta m^2
- 1\right] =0
\ ,
\label{3011}
\ee
where $W'=\d W/\d r$.  
Since the expression inside the square brackets cannot vanish, one must have
\be
\partial _ {\theta}{\mathit{\Theta}}
+i\,\frac{\partial _ {\phi}{\mathit{\Theta}}}{\sin\theta}
=
0
\ ,
\label{eqTheta}
\ee
and the solution for ${\mathit{\Theta}} $ will therefore give no contribution to the tunnelling rate.
Next, on substituting Eq.~(\ref{3010}) with Eq.~\eqref{eqTheta} into Eqs.~(\ref{306}) and~(\ref{307}),
and again factoring out ${{\psi} }$ and ${{\mathring\psi} }$, we obtain 
\be
{\psi}_0+{\psi}_2\left(W'\right)^{2}+{\psi}_4\left(W'\right)^{4}+{\psi}_6\left(W'\right)^{6}
=
0
\ ,
\label{3014}
\ee
where
\begin{subequations}
\ba
\psi_0
&=&
-\left[m^{2}\,A+\left(e\,{\rm A}_t\right)^2\right]
\left(1-\beta\, m^{2}\right)^2
-\omega^2+2\,\omega\, e\,{\rm A}_t
\left(1-\beta\, m^{2}\right)
\ ,
\\
\psi_2
&=&
\beta\, B
\left\{2\, e\,{\rm A}_t
\left[e\,{\rm A}_t
\left(1-\beta\, m^2\right)
-\omega\right]
+
A\left(1-\beta^2\,m^4\right)
\right\}
\ ,
\\
\psi_4
&=&
-\beta\, B^2
\left[\beta \left(e\,{\rm A}_t\right)^2
+\beta\, A
\left(2-\beta\,m^{2}\right)
\right]
\ ,
\\
\psi_6
&=&
\beta^{2}\,B^{3}\,A
\ .
\label{3015}
\ea
\end{subequations}
Solving
Eq.~(\ref{3014}) on the event horizon yields the imaginary part of the action, 
\be
{\rm Im}\, W_\pm (r)
=
\pm \frac{\pi}{4}\,
\frac{r_+^2\,\omega
\left(1+ \beta\,\Xi\right)}
{r_+ - a\,r_-}
\ ,
\ee
where
\begin{subequations}
\begin{eqnarray}
a
&=&
\frac{4\,M\left(16\, M^3/\mpl^3+M/\mpl+\ell/\lp\right)}{\mpl\left(3\,M/\mpl-2\,\ell/\lp\right)\left(M/\mpl+\ell/\lp\right)}
\ ,
\label{a}
\\
\Xi
&=&
\frac{3}{2}\,m^2
+{\frac{e\,m^2\,{\rm A}_t}{\omega-e\,{\rm A}_t}
-\frac{2\,e^2\,{\rm A}_t\left(7\,M/\mpl-2\,\ell/\lp\right)}{M/\mpl+2\,\ell/\lp}}
+\frac{2\,M\,\omega}{\left(M/\mpl+2\,\ell/\lp\right)}
\ . 
\label{3017}
\end{eqnarray}
\end{subequations}
Thus, the tunnelling rate of fermions reads 
\be
\Gamma
\simeq
\frac{\exp\!\left(- 2\,\mathrm{Im} {{\mathit{\Theta}}}-2\,\mathrm{Im} W_+\right)}
{\exp\!\left(-2\,\mathrm{Im} {{\mathit{\Theta}}}-2\,\mathrm{Im} W_-\right)}
\simeq
\exp\!\left\{-\frac{8\,\pi\,M^2\left(1 + \beta\,\Xi\right)\omega}
{\mpl^3\left(M/\mpl+\ell/\lp\right)}
\right\}
\ ,
\label{3018}
\ee
in which we just kept the leading order for $\ell\lesssim \lp\,M/\mpl$ in the coefficient~\eqref{a}.
From now on, we shall only consider neutral fermions ($e=0$), so that
\be
\beta\,\Xi
=
\beta_0
\left[
\frac{3\,m^2}{2\,\mpl^2}
+\frac{2\,M\,\omega}{\mpl^2\left(M/\mpl+2\,\ell/\lp\right)}
\right]
\ ,
\ee
and the corresponding rate $\Gamma$ is plotted in Figs.~\ref{b1}-\ref{bs3}. 
\par
The rate~\eqref{3018} can be written as the Boltzmann-like factor
$\Gamma=\exp\left(-{\omega}/{T}\right)$, where
\be
T
=
\frac{\mpl^2\left(M+\,\ell_0\,\mpl\right)}{8\,\pi\, M^2\left(1+\beta\,\Xi\right)}
\ ,
\label{3019}
\ee
and $\ell_0=\ell/\lp$.
We then observe that, for $M\gg\mpl$ (and $\omega\sim\mpl^2/M\ll M$), the above expression
can be approximated as
\be
T
\simeq
T_0
\left(
1
-2\,\beta_0\,\frac{\mpl}{M}
\right)
\label{Tlarge}
\ee
where
\be
T_0
=
\frac{\hbar}{4\,\pi} \left[\sqrt{A'(r)\,B'(r)}\right]_{r=r_+}
=
\frac{\mpl^2}{8\,\pi\,M}\,\sqrt{1+\frac{2\,\ell_0\,\mpl}{M}}
\simeq 
\frac{\mpl^2}{8\,\pi\,M}\,\left(1+\ell_0\,\frac{\mpl}{M}\right)
\label{T0}
\ee
is the Hawking temperature of the MGD black hole obtained with the tunnelling method~\cite{k0}
(and reduces to the standard Hawking expression for $\ell_0=0$).
The tunnelling rate~\eqref{3018} therefore reproduces the Hawking result for large size black holes
with a mass sufficiently large such that both the GUP correction (proportional to $\beta_0$) in Eq.~\eqref{Tlarge}
and the MGD correction (proportional to $\ell_0$) in Eq.~\eqref{T0} remain negligible. 
\par
For black hole mass approaching the Planck scale, the dependence of $\Xi$ on $\omega\sim M\sim \mpl$ cannot be
neglected, and one cannot just consider the emission at a fixed temperature for all frequencies.
We can still assume the fermion mass $m\simeq 0$, since we consider $M$ at least of the order of the
Planck scale, which is about 19 orders of magnitude heavier than the heaviest fundamental fermions ever observed. 
We then regard Hawking particles with given energy $\omega$ in the angular mode $l$. 
According to Eq.~\eqref{3018}, these particles will be emitted with a probability approximately given by the rate
${\Gamma(\omega)}=\exp\left[-{\omega}/{T(\omega)}\right]$ multiplied by the probability for the black hole
to absorb such particles.
Since the average number of fermions per mode, 
\be
n_{l}(\omega)
=
\frac{1}{e^{\omega/T}+1}
=
\frac{\Gamma(\omega)}{\Gamma(\omega)+1}
\ ,
\ee
is related to the average emission rate $\dot n_l(\omega)=\d n_l(\omega)/\d t$
by ${2\,\pi\,\hbar}\,\dot n_l(\omega)=n_l(\omega)\, \d\omega$~\cite{Page:1976df}, 
the total luminosity of the black hole can be obtained by multiplying by $\omega$ and summing over the modes, that is 
\be
L
=
{\displaystyle\sum\limits_{l=0}}
\left(  2\,l+1\right)
\int\left\vert T_{l}\left(  \omega\right)  \right\vert^{2}
n_l(\omega)\,\frac{\omega\,\d\omega}{2\,\pi\,\hbar}
\ , 
\ee
where $T_{l}\left(  \omega\right)$ are the grey-body factors. 
In the geometric optics regime, $\omega\gg \mpl^2/M$, the sum over the discrete angular modes $l$
can be approximated with an integral and the luminosity then reads
\be
L
=
\frac{T^{4}\,M^{2}}{2\,\pi\,\mpl^5\,\lp}
\int_0^{\infty}\left(\frac{\omega}{T}\right)^{3}\d\!\left(\frac{\omega}{T}\right)
\int_0^{\gamma}
n
\left[
\left(\frac{\omega}{T}\right)
\left(  1+\frac{l\left(  l+1\right)\mpl^{6}}{128\,\pi^2\, T\,\omega\, {M^4}}\right)
\right]
\d\!\left(\frac{l\,(l+1)\,\mpl^4}{M^2\,\omega^2}\right)
\ ,
\label{lum222}
\ee
where the upper integration limit $\gamma$ arises from studying wave scattering by the black hole.
In this respect, a black hole can be viewed as a black sphere, with an upper bound on the angular modes
that can be absorbed given by $l\,(l + 1)\,\mpl^4\lesssim 27\,M^2\,\omega^2$~\cite{Mu:2015qta}. 
Since modes with $l$ exceeding this bound, for a given $\omega$, are not absorbed by the black hole,
they cannot be emitted either.
\par
{\color{black}
The flux~\eqref{lum222} can indeed be computed with the complete expression for $T$ in Eq.~\eqref{3019},
but turns out to be very cumbersome.
For $M\gg\mpl$, the leading behaviour is given by 
{\color{black}
\be
L
\simeq
\frac{\pi \,\mpl^3}{8\,\lp\,M^2}
\ ,
\ee
and the evaporation rate $\dot M\simeq -L$ reproduces the expected Hawking expression for large
black holes, in agreement with the above analysis of the temperature $T$.}
\par
On the other hand, for $M\simeq\mpl$, the flux is given by
\be
\label{flux}
L(M,\ell_0,\beta_0,m)
=
\frac{\pi \,\mpl^3}{8\,\lp\,M^2}
+\beta_0\,\frac{\mpl}{\lp}
\left(\frac{\pi \,\mpl^2}{35\,M^2}-\frac{M-2\,\mpl}{8\,\mpl}\right)
-\ell_0\,\frac{\mpl}{\lp}\left(\frac{M^2}{\mpl^2}+F\right)
\ ,
\ee
where $F$ contains higher powers of $M$ and of the parameters $\beta_0$ and $\ell_0$, but is rather
involved and we will just display its expansion to order $M^5/\mpl^5$ for completeness below.
The evaporation rate $\dot M\simeq -L$ with the flux~\eqref{flux} again reproduces the Hawking
expression for $\beta_0=\ell_0=0$, but can vanish for a finite value of $M=M_{\rm c}$ otherwise.}
The explicit expression of $M_{\rm c}$ is again very complicated and we shall just estimate it in
special regimes of the parameters $\ell_0$ and $\beta_0$.  
For $\ell_0=0$ and $\beta_0>0$ the second term in Eq.~\eqref{flux} can be negative and compensate
for the Hawking contribution.
In particular, for $\ell_0=0$ and $0<\beta_0\ll 1$, the flux vanishes for
\be
M_{\rm c}(\beta_0)
\sim
\beta_0^{-1/3}\,\mpl
\ ,
\label{critical1}
\ee
which can be rather large.
 Of course, $\beta_0\ll 1$ means that the length $\sqrt{\beta}\ll \lp$, and it is therefore more sensible to consider $\beta_0\simeq 1$ for which we obtain
\be
M_{\rm c}
\sim
2.6\,\mpl
\ .
\ee
The above result does not change significantly for $\ell_0=0$ and $\beta_0\gg 1$ (we recall that current bounds on $\beta_0$ are rather
large~\cite{Das:2008kaa,Scardigli:2014qka}), {since} the critical mass in this limit is  given by 
\be
M_{\rm c}
\sim
2.2\,\mpl
\ .
\label{critical2}
\ee
{Conversely}, when $\beta_0=0$ and $0<\ell_0\ll 1$ (so that $F$ is negligible), the third term in Eq.~\eqref{flux} 
leads to the flux vanishing at a critical mass given by
\be
\label{critical3}
M_{\rm c}(\ell_0)
\sim
\ell_0^{-1/4}\, \mpl
\ , 
\ee
which could again be fairly large. As for $\beta_0$, it is physically more sensible to consider $\ell_0\sim 1$, for which the critical mass can be estimated numerically and is given by
\be
M_{\rm c}(\ell_0)
\sim
2.1\, \mpl
\ . 
\ee
\par
We mentioned above that higher order (in $M/\mpl$) corrections to the flux are contained in the function
\ba
F(M,\ell_0,\beta_0,m)
&\simeq&
\frac{315\,\pi\,\beta_0^2\,\ell_0\,\mpl^2}{M^2}
\!+\!\frac{M^2\,\ell_0}{\mpl^2}
\left[\frac{2\,\pi}{5}\left(3\,\beta_0\,\frac{m^2}{\mpl^2}\right)
-\frac{\ell_0}{16}\right]
\nonumber
\\
&&
\!+\!\frac{M^3\,\ell_0^2}{160\,\mpl^3}
\left[5-32\, \pi\,\ell_0^3\left(1\!+\!3\, \beta_0\,\frac{m^2}{\mpl^2}\right)\right]
\nonumber
\\
&&
\!+\!\frac{M^4\,\ell_0}{\mpl^4}
\left[
\frac{\ell_0^2 }{10}
\left(1\!+\!3\, \beta_0\,\frac{m^2}{\mpl^2}\right)
-
\left(1\!+\!3\, \beta_0\,\frac{m^2}{\mpl^2}\right)^2
-
\frac{\ell_0^6}{\pi}
\right]
\nonumber
\\
&&
\!+\!\frac{M^5\,\ell_0}{13440\,\mpl^5}
\left[105
-
672\, \pi\,\ell_0^3
\left(1\!+\!3\, \beta_0\,\frac{m^2}{\mpl^2}\right)
\!+\!
2240\,\pi\,\ell_0^7
\left(1\!+\!3\, \beta_0\,\frac{m^2}{\mpl^2}\right)^2
\right]
\ ,
\quad
\label{flux1}
\ea
which will affect the precise value of the critical mass $M_{\rm c}$.
Let us note once more that the correction~(\ref{flux1}) vanishes for $\ell_0=\beta_0=0$, as expected.
Moreover, typical fermions emitted by black holes will satisfy the bound $\beta_0\,{m^2}\lesssim 10^{-16}\,{\mpl^2}$
and, since all terms of order $\mathcal{O}(M^6)$ are multiplied by factors of $\beta_0^2\,{m^4}/{\mpl^4}$,
the expression given above for the function $F$ should be quite accurate for quantum black holes.
\section{Concluding remarks}
\label{SIV}
The tunnelling method was here employed as a dynamical model of Hawking radiation in order to compare
the effects stemming from the MGD of the background black hole metric with those described by a GUP
applied to the quantum dynamics of the emitted particles. 
It is important to stress that both the MGD and GUP corrections are of (quantum) gravitational origin, and such
a comparison could be useful for probing {{sectors of the semiclassical approach of quantum gravity}}.\par
Our calculations were based on the Hamilton-Jacobi method, which employs the WKB approximation
in order to solve the generalised Dirac equation, on a MGD background geometry.
In particular, we found that the total flux of fermions emitted by the black hole can vanish for a critical
mass~(\ref{critical1})-(\ref{critical3}), depending on the relative strength of the MGD parameter $\ell$
and the GUP parameter $\beta$. 
Hence, black hole remnants, with a mass determined by these two length scales could exist,
say in the range between the critical masses~(\ref{critical2}) and~(\ref{critical1}) or \eqref{critical3},
if also boson emission is suppressed in the same regime. \textcolor{black}{These remnants are also stable under small linear perturbations. 
In summary, the Hawking temperature of the MGD black hole is shown to be corrected by GUP effects and the
black hole evaporation is in general reduced by quantum effects of gravity.}
\acknowledgments
R.C.~is partially supported by the INFN grant FLAG.
The work of P.N.~has been supported by the project ``Evaporation of the microscopic black holes'' of the German
Research Foundation (DFG) under the grant NI~1282/2-2.
R.dR.~is grateful to CNPq (Grant No. 303293/2015-2),
and to FAPESP (Grant No. 2017/18897-8) and to INFN, for partial financial support.

\end{document}